\documentclass[floatfix,aps,superscriptaddress,twocolumn,pra]{revtex4-2}
\usepackage{amssymb,amsmath,amsfonts,mathrsfs,bm, bbm, graphicx, epsfig, epstopdf,soul,physics,cancel,mathtools}
\usepackage{tabularray}
\usepackage[T1]{fontenc}
\usepackage[version=3]{mhchem}
\usepackage{braket}
\usepackage{hyperref}
\usepackage{color}
\usepackage{csquotes}
\usepackage{float}
\usepackage{siunitx}
\usepackage[normalem]{ulem}
\usepackage{gensymb}

\usepackage{xcolor}

\begin{document}
\title{A global quantum network with ground-based single-atom memories in optical cavities and satellite links}
\author{Jia-Wei Ji}
\email{quantum.jiawei.ji@gmail.com}
\affiliation{Institute for Quantum Science and Technology, and Department of Physics \& Astronomy, University of Calgary, 2500 University Drive NW, Calgary, Alberta T2N 1N4, Canada}

\author{Shinichi Sunami}
\affiliation{Nanofiber Quantum Technologies, Inc. (NanoQT),
1-22-3 Nishiwaseda, Shinjuku-ku, Tokyo 169-0051, Japan}

\affiliation{Clarendon Laboratory, University of Oxford, Oxford OX1 3PU, United Kingdom}

\author{Seigo Kikura}
\affiliation{Nanofiber Quantum Technologies, Inc. (NanoQT),
1-22-3 Nishiwaseda, Shinjuku-ku, Tokyo 169-0051, Japan}

\author{Akihisa Goban}
\email{akihisa.goban@nano-qt.com}
\affiliation{Nanofiber Quantum Technologies, Inc. (NanoQT),
1-22-3 Nishiwaseda, Shinjuku-ku, Tokyo 169-0051, Japan}

\author{Christoph Simon}
\email{christoph.simon@gmail.com}
\affiliation{Institute for Quantum Science and Technology, and Department of Physics \& Astronomy, University of Calgary, 2500 University Drive NW, Calgary, Alberta T2N 1N4, Canada}

\begin{abstract}
The realization of a global quantum network holds the potential to enable groundbreaking applications such as secure quantum communication and blind quantum computing. However, building such a network remains a formidable challenge, primarily due to photon loss in optical fibers. In this work, we propose a quantum repeater architecture for distributing entanglement over intercontinental distances by leveraging low-Earth-orbit satellites equipped with spontaneous parametric down-conversion (SPDC) photon-pair sources and ground stations utilizing single-atom memories in optical cavities and single-photon detectors to implement the cavity-assisted photon scattering (CAPS) gates for high-fidelity entanglement mapping. The efficient entanglement swapping is achieved by performing high-fidelity Rydberg gates and readouts. We evaluate the entanglement distribution rates and fidelities by analyzing several key imperfections, including time-dependent two-photon transmission and time-dependent pair fidelity, for various satellite heights and ground station distances. We also investigate the impact of pair source fidelity and spin decoherence rate on the repeater performance. Furthermore, we introduce a spatial-frequency multiplexing strategy within this architecture to enhance the design's performance. Finally, we discuss in detail the practical implementation of this architecture. Our results show that this architecture enables entanglement distribution over intercontinental distances. For example, it can distribute over 10000 pairs per flyby over 10000 km with a fidelity above $90\%$, surpassing the capabilities of terrestrial quantum repeaters.  
\end{abstract}

\maketitle

\section{Introduction}\label{sec:introduction}
The development of a global quantum network promises a range of exciting applications, such as secure communication \cite{RevModPhys.74.145}, blind quantum computing \cite{Barz303}, private database searches \cite{PhysRevA.83.022301}, and ultimately a quantum internet that links quantum computers and other quantum information systems \cite{Kimble2008,Simon2017,Wehnereaam9288,RevModPhys.95.045006}. In these networks, photons serve as the carriers of information for establishing long-range connections. However, transmission losses significantly hinder their ability to connect distant locations. Unlike in classical communication, photon loss cannot be solved by amplification in quantum communication because the no-cloning theorem prohibits the perfect duplication of unknown quantum states \cite{Wootters1982}. To address this challenge, quantum repeaters have been introduced, which divide the overall connection into shorter, more manageable segments \cite{PhysRevLett.81.5932} to create and store entanglement. Then, entanglement swapping is performed to propagate the entanglement from end to end. So far, numerous theoretical approaches to quantum repeaters based on different systems have been proposed, accompanied by proof-of-principle experimental demonstrations. This includes atomic ensembles \cite{Duan2001,RevModPhys.83.33,Liu2021,Lago-Rivera2021,Liu2024}, trapped ions \cite{RevModPhys.82.1209,PhysRevLett.130.050803}, superconducting qubits \cite{Kurpiers2018,Kumar_2019,Zhong2019,Zhong2021}, single rare earth ions \cite{KimiaeeAsadi2018quantumrepeaters,Asadi_2020,Ourari2023}, color centers in diamond \cite{Hensen2015,Humphreys2018,Bhaskar2020,doi:10.1126/science.abg1919,Knaut2024}, and neutral atoms \cite{doi:10.1126/science.1221856,RevModPhys.87.1379,PhysRevLett.124.063602,PhysRevLett.126.130502,doi:10.1126/science.abe3150}. Although these repeater schemes show great promise in beating direct transmission, the achievable distance is typically limited to a few thousand kilometers even with multiplexing. The memory-less quantum repeaters such as the one-way quantum repeaters \cite{Muralidharan2016,PhysRevX.10.021071,RevModPhys.95.045006} and the VBG (vacuum beam guide) approach \cite{PhysRevLett.133.020801} have been proposed to overcome the limits of memory-based quantum repeaters, holding promise in distributing entanglement at an intercontinental scale, but they require complex quantum error correction and demand significant investment in very costly infrastructure for deployment.

As opposed to the exponential loss in optical fibers, which is a fundamental challenge in terrestrial quantum repeaters, the satellite links mainly suffer from diffraction in the vacuum of space, which has an inverse square scaling, thus providing a promising way to go beyond the limits of ground-based quantum repeaters. Since the launch of the Micius quantum satellite, several groundbreaking experiments in satellite QKD (quantum key distribution) and entanglement distribution have been demonstrated \cite{Liao2017,Ren2017,doi:10.1126/science.aan3211,PhysRevLett.120.030501,Yin2020,Chen2021,RevModPhys.94.035001,Li2025}. These demonstrations either used direct transmission from low-Earth-orbit (LEO) satellites or trusted nodes, which are still far from achieving global-scale secure quantum communications. An interesting way to build a global quantum network is to combine the idea of quantum repeaters and satellite links, and two possible solutions are proposed, with one using ground-based repeaters connected by satellites \cite{PhysRevA.91.052325} and the other fully space-based using satellites equipped with quantum memories \cite{Gundogan2021}. Moreover, a memory-less satellite protocol has also been investigated to distribute entanglement at global distances, known as the satellite-relayed scheme \cite{PhysRevApplied.20.024048}. This requires a carefully aligned chain of satellites with ``satellite lenses" to eliminate diffraction losses, but potentially introducing other kinds of loss.

\begin{figure*}
\centering
  \includegraphics[width=0.99\textwidth]{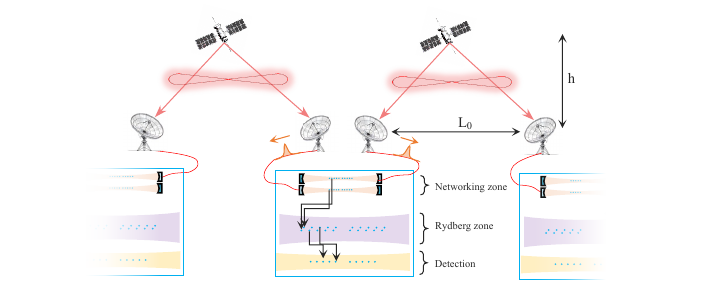}
  \caption{Proposed satellite-assisted quantum repeater architecture. Here, we show two links to illustrate our repeater scheme. $L_0$ stands for the elementary link length and $h$ is the satellite hight. The LEO satellites are equipped with SPDC photon-pair sources. The first step is to generate atom-atom entanglement in each elementary link by mapping the photon-photon entanglement. This is achieved by performing the cavity-assisted photon scattering (CAPS) gate between the photon and the atom in the cavity and detecting the photon in each node. Once both links are established, the second step is to implement entanglement swapping by performing the Rydberg gate between the atoms, followed by local single-qubit rotations and state-selective readouts in the X basis. 
  }
  \label{Figure 1}
  \end{figure*}

Here, we propose to use single atoms as memories in the context of quantum repeaters with satellites. The tweezer-trapped single atoms exhibit long coherence times \cite{Young2020, PhysRevX.12.021028,Bluvstein2022,Barnes2022}, making it possible to perform various coherent operations. Most importantly, high-fidelity Rydberg gates between two atoms have been demonstrated \cite{PhysRevLett.123.170503,PhysRevA.105.042430,PhysRevX.12.021028,Evered2023}, enabling deterministic operations for entanglement swapping. In addition, the cavity-assisted photon scattering (CAPS) gate between photon and atom has also been demonstrated \cite{Hacker2016,PhysRevX.8.011018}, allowing near-deterministic and high-fidelity operations for entanglement mapping from photons to atoms. These atoms possess optically active transitions in a wide range of  wavelengths, including the telecom band, providing the possibility of interfacing with the SPDC source on the satellite and allowing for integration with the terrestrial quantum repeaters \cite{PhysRevResearch.3.043154}.

The recent proposal \cite{satelliteassistedquantumcommunicationsingle} focuses on space-borne atomic memories in cavities for global entanglement distribution. The main challenge remains navigating the complexities of operating such a system in space. In contrast to this work, we keep the main complexity on the ground by considering ground stations with single-atom memories in optical cavities and detectors to store the entanglement of the entangled photon pairs from satellites \cite{PhysRevA.91.052325}. The CAPS gate is performed between the incoming photon and the atom in the cavity, and the detector is used for heralding to map the photon-photon entanglement to the atom-atom entanglement with high fidelity and a high success probability. Then, Rydberg gates are performed between the successful atom pairs to achieve deterministic entanglement swapping for propagating the entanglement to the end nodes. We analyze the performance of the repeater by incorporating various key factors, including the time-dependent two-photon transmission, time-dependent photon-pair fidelity, and spin decoherence. Specifically, we investigate the effects of satellite altitude, pair source fidelity, and spin decoherence rate on both entanglement generation rates and fidelities. Moreover, we introduce a spatial and frequency multiplexing strategy to improve the repeater's performance, which allows for feasible rates and fidelities of entanglement distributed at global distances.

This paper is organized as follows. In Sec. \hyperref[sec:quantum archi]{II}, we describe the quantum network architecture by focusing on the satellite-to-ground optical links and entanglement mapping via the CAPS gate. Sec. \hyperref[sec:rates]{III} discusses the entanglement distribution rates and fidelities and explores the impact of various important parameters on the repeater performance. Sec. \hyperref[sec:implementation]{IV} gives details on system implementation. We conclude and provide an outlook in Sec. \hyperref[sec:conclusion]{V}.

\section{The quantum network architecture} 
\label{sec:quantum archi}
The diagram in Fig.~\ref{Figure 1} illustrates the basic steps and components for building a global quantum network based on single-atom memories in cavities and satellite links. This architecture features two basic steps: the entanglement generation between remote atoms in cavities and entanglement swapping between local atoms to propagate the entanglement.  

This section is dedicated to the three basic components of our proposed architecture. The first one is the optical downlinks from the satellite to the ground stations, where we quantify the imperfections in this process and take the dynamics of the satellite into account. Then, we show how photon-photon entanglement can be mapped to atom-atom entanglement via the CAPS gate, and quantify their performance. In the entanglement swapping, we discuss how the swapping can be implemented in our system to propagate the entanglement. The details of implementation are presented in Sec. \ref{sec:implementation}.

\subsection*{Repeater operation}
\label{subsec:atom}

\subsubsection{Quantum optical downlinks}
\label{sec:downlink}

In our architecture, we focus on space-to-ground links, also called downlinks. For each elementary link, the LEO satellite is equipped with an SPDC source that emits polarization-entangled photon pairs with their quantum state given by:

\begin{equation}
\begin{aligned}
\rho_s=F_s\ket{\Psi^+}\bra{\Psi^+}+\frac{1-F_s}{3}&(\ket{\Phi^-}\bra{\Phi^-}+\ket{\Phi^+}\bra{\Phi^+}\\
&+\ket{\Psi^-}\bra{\Psi^-}),  
\end{aligned}
\end{equation}
where $\ket{\Psi^\pm}=\frac{1}{\sqrt{2}}(\ket{\mathrm{HV}}_p\pm\ket{\mathrm{VH}}_p)$ and $\ket{\Phi^\pm}=\frac{1}{\sqrt{2}}(\ket{\mathrm{HH}}_p\pm\ket{\mathrm{VV}}_p)$ with $\mathrm{H(V)}$ standing for the horizontal(vertical) polarized state, and $F_s$ is the source fidelity. This type of source can be both spatially and spectrally multiplexed, greatly boosting its performance, and technical details are discussed in the implementation section \ref{sec:implementation}. When reaching the ground stations, the emitted photon pairs typically suffer from three main losses: diffraction loss, atmospheric absorption, and pointing loss. The polarization of incoming photons is continuously adjusted based on real-time polarization analysis of the reference beams during the flyby \cite{RevModPhys.94.035001} to ensure the polarization matching between the satellite and ground stations. Assuming the beam takes the Gaussian profile, the diffraction loss is given by \cite{Gundogan2021}:
\begin{equation}
 \eta_{\text{diff}}=1-\exp(-\frac{D^2_\text{R}}{2\omega^2_d}),   
\end{equation}
where $\omega_d=\sqrt{\omega^2_ 0(1+(\frac{\lambda d}{\pi\omega^2_ 0})^2)}$ is the diffracting beam waist that depends on the photon wavelength $\lambda$, original waist $\omega_0$, and the distance $d$ between the transmitter and receiver. $D_\text{R}$ stands for the receiver radius (telescope) in the ground station. The distance $d$ is a function of time as the satellite moves in a polar orbit between the two ground stations.  

The atmospheric transmittance is given by \cite{Khatri2021}:
\begin{equation}
\eta_{\text{atm}}=(\eta_{\text{zenith}})^{\sec\theta}, 
\end{equation}
where $\eta_{\text{zenith}}$ is the transmittance at zenith, and $\theta$ is the zenith angle, which is defined as \cite{Khatri2021}:
\begin{equation}
\cos\theta=\frac{h}{d}-\frac{1}{2}\frac{d^2-h^2}{R_\text{E} d}, 
\label{eq:zenith}
\end{equation} 
where $R_E\approx 6378$ km is the Earth's radius, and this is only valid for small zenith angles, i.e., $0<\theta<\pi/2$ \cite{Khatri2021}. $h$ is the satellite height. The maximum zenith angle is determined by the lowest elevation angle of the telescope in the ground station, below which we lose the line of sight. The transmittance at zenith is determined by photon wavelength, and the specific values as a function of wavelength are given in \cite{doi:10.1139/cjp-2023-0190}. 

\begin{figure}
\centering
  \includegraphics[scale=0.55]{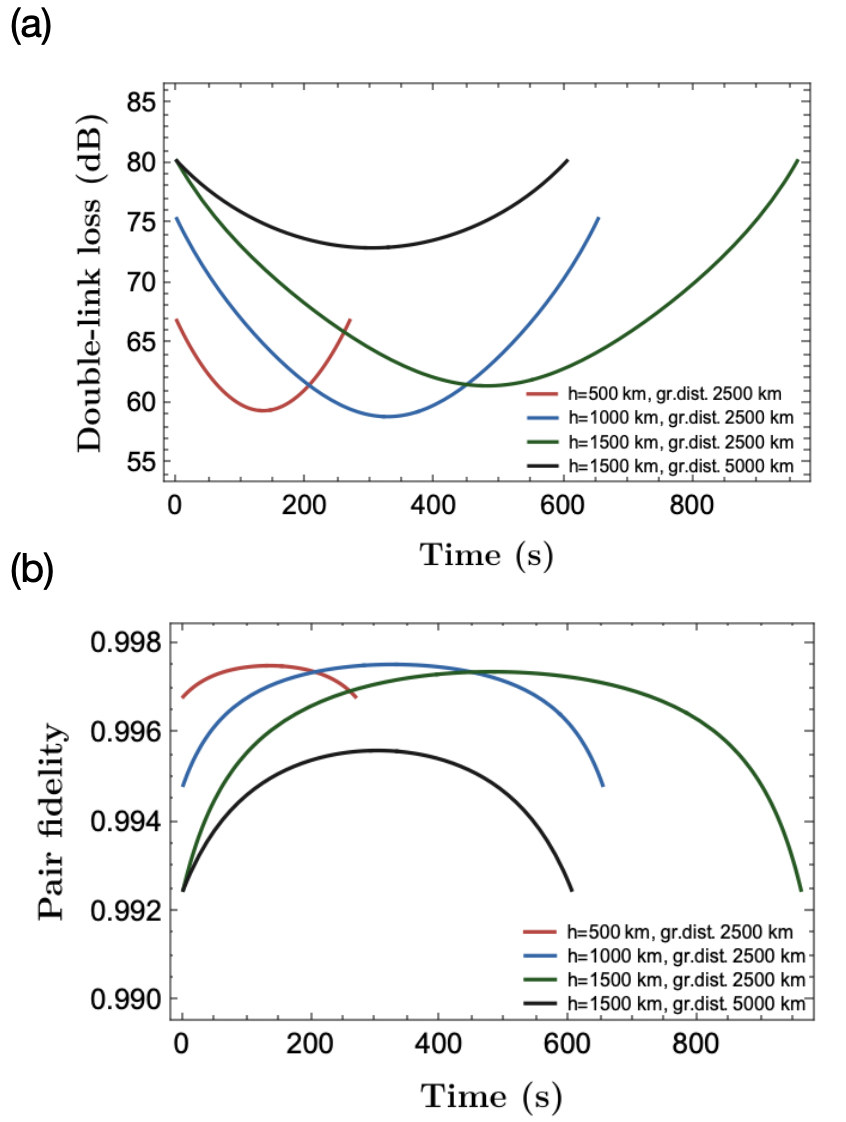}
  \caption{(a) Two-photon transmission from the satellite as a function of time. The time window is the entire flyby time for all plots. The curves correspond to the cases with different satellite heights and ground station distances. Both the height and the ground station distance affect the flyby time. (b) Pair fidelity as a function of time. The curves show the evolution of the pair fidelity during the flyby for various satellite heights and ground distances. Table \ref{table: parameters} shows the other parameters used in this figure.}  
  \label{tptrans}
  \end{figure}

For a single flyby, the Earth's rotation is negligible. Then, the time-dependent distance $d$ is given by:
\begin{equation}
\begin{aligned}
d^2(t)=&R^2_E+(R_E+h)^2\\
-&2R_E (R_E+h)\cos\frac{L_0}{2R_E}\cos\omega(t_ 0-t), 
\end{aligned}
\label{eq:dis}
\end{equation}
where $L_0$ is the elementary link length, and $\omega=\sqrt{GM_E/(h+R_E)^3}$ is the angular speed of the satellite with $G$ being the gravitational constant and $M_E$ being the Earth's mass. $2t_0=T_{\text{FB}}$ is the flyby time. When $t=t_0$, the satellite is exactly over the head of the midpoint between the two ground stations. The time $t_0$ is determined by the maximum zenith angle with a given elementary link length $L_0$. When the link length passes a certain value, the flyby time becomes zero. The detailed derivations of the time-dependent distance $d(t)$ and the time $t_0$ can be found in Appendix \ref{app:dynamics}.


The pointing loss occurs when tracking a high-speed satellite from the ground, affecting the establishment of the link. This error can be quantified using the following equation \cite{RevModPhys.94.035001}:
\begin{equation}
\eta_{\text{point}}=\frac{\Delta\theta^2}{\Delta\theta^2+4\sigma^2_{\text{point}}},    
\end{equation}
where $\sigma_{\text{point}}$ is the Gaussian pointing probability density standard deviation. $\Delta\theta=4M^2\lambda/(\pi\omega_0)$ is the beam divergent angle with $M^2$ being the beam quality factor ($M^2=1$ for the perfect Gaussian beam).

In a downlink configuration, the effect of air turbulence on wavefront distortion can be neglected, as the propagated beam traverses the atmosphere only at the end of its path. However, turbulence can cause beam wander, reducing coupling efficiency at the receiver \cite{doi:10.1139/cjp-2023-0190}. Therefore, we include the inefficiency due to this effect and other coupling inefficiencies in one efficiency and call it the system coupling efficiency $\eta_{\text{cpl}}$. Due to the symmetry, the time-dependent two-photon transmission can be written as:
\begin{equation}
\eta^{(2)}_{\text{tr}}(t)=\eta^2_{\text{tr}}(t)=\eta^2_ {\text{dif}}(t)\eta^2_ {\text{atm}}(t)\eta^2_{\text{point}}\eta^2_{\text{cpl}},
\end{equation}
where $\eta^{(2)}_{\text{tr}}(t)$ is the two-photon transmission. Fig. \ref{tptrans}(a) shows the two-photon transmission as a function of time for different satellite heights and ground station distances.

The above imperfections cause photon loss, thus only impacting the final distribution rates. However, during the transmission from space to the ground, the background photons can cause noise, which degrades the photon pair fidelity. The fidelity of the photon pair when reaching the ground is given by \cite{Khatri2021}:
\begin{equation}
F_{\text{pair}}(t)=\frac{1}{4}\left\{1+\frac{4F_s-1}{(1+\frac{\bar{n}}{\eta_{\text{tr}}(t)})^2}\right\},  
\label{eq:pairfid}
\end{equation}
where $F_s$ is the source fidelity of the photon pair and $\bar{n}$ is the mean background photon number, which is given by \cite{Khatri2021}:
\begin{equation}
 \bar{n}=\frac{H\Omega_{\text{fov}}\pi(\frac{D_{\text{R}}}{2})^2\Delta\lambda\Delta T}{hc/\lambda},   
\end{equation}
where $H$ is the spectral irradiance of the sky (brightness), $\Omega_{\text{fov}}$ is the field of view, $\Delta\lambda$ is the filter bandwidth, $\Delta T$ is the coincidence time window, and the denominator $hc/\lambda$ is the photon energy with $h$ the Planck's constant and $c$ the speed of light. The values of these parameters are given in Table \ref{table: parameters}. Fig. \ref{tptrans}(b) shows the pair fidelity as a function of time during the flyby for various satellite heights and ground station distances. Although the sky brightness $H$ is assumed to correspond to a moonless clear night, this condition could be relaxed by having a smaller field of view and a smaller telescope size.

\subsubsection{Entanglement mapping via the CAPS gate}
\label{sec:entmap}

\begin{figure}
    \centering
  \includegraphics[width=0.9\linewidth]{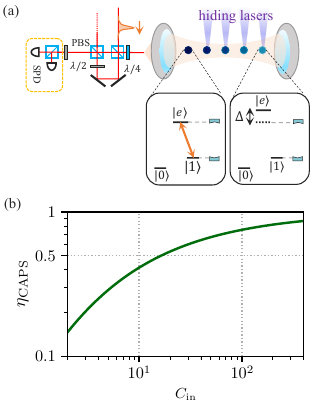}
  \caption{Entanglement mapping via photon reflection from the cavity. 
  (a) Schematic of time-multiplexed CAPS gate: the cavity comprises two mirrors (planes at both ends), between which is the cavity mode (orange).
  Optical tweezer array traps individual atoms (circles), and one atom is coupled to the cavity mode while the other atoms are detuned from the cavity to avoid crosstalk.
  (b) Memory loading success probability as a function of the internal cooperativity $C_\mathrm{in}$. The infidelity arising from reflectivity mismatch between two nodes can be compensated by inserting controlled attenuation in the V-polarized path.   
  }  
  \label{fig:duan_kimble}
  \end{figure}

The entangled photons received by the ground station are directed to nearby repeater devices, each consisting of an optical cavity coupled to a neutral-atom array (see Fig.~\ref{Figure 1}).
These photons are delivered through optical fibers and routed to individual atoms within the cavity mode, where only one atom is resonantly coupled at a time, while the others are detuned via an AC Stark shift, as illustrated in Fig.~\ref{fig:duan_kimble}(a).
In this configuration, the atom-cavity system implements the CAPS-gate protocol, enabling heralded transfer of photonic polarization states onto atomic qubits.

The process begins with a polarization-entangled photon pair in the state $(\ket{\mathrm{HV}}_p + \ket{\mathrm{VH}}_p)/\sqrt{2}$, shared between two distant nodes.
At each node, the incoming photon is sent to a polarizing beamsplitter (PBS) that splits the two polarization components along separate paths: the V-polarized component reflects from the cavity mirror, whereas the H-polarized component reflects from a reference mirror.
A set of waveplates is adjusted so that the two paths are recombined and the photon exits the opposite port of the PBS. 
Inside the cavity, the photon interacts with an atom, initialized in the superposition state $\ket{+}_a = (\ket{0}_a + \ket{1}_a)/\sqrt{2}$, where only the $\ket{1}_a \leftrightarrow \ket{e}_a$ transition is resonant with the cavity. 
When the cavity, the photon, and the atomic transition are all on resonance, the reflected light acquires a $\pi$ phase shift if the atom is in $\ket{0}_a$.
This conditional phase shift realizes a controlled-phase (CZ) gate between the atomic and photonic qubits \cite{Duan2004}. 
A subsequent Hadamard gate on the photon, followed by an $X$-basis measurement, teleports its state onto the atom; executing this sequence at both nodes maps the original photonic entanglement onto a remote atomic Bell pair.

The fidelity of this entanglement mapping process is set by the CAPS gate performance at each node.
The leading sources of infidelity are amplitude and temporal-mode mismatches between the H- and V-polarized photons, which depend on the atom-cavity parameters $(g, \kappa_\mathrm{ex}, \kappa_\mathrm{in},\gamma)$: the atom-cavity coupling $g$, the external coupling rate $\kappa_\mathrm{ex}$, the internal loss rate $\kappa_\mathrm{in}$, and the atomic decay rate $\gamma$. 
The cavity reflectivities for the uncoupled and coupled atomic states are given by $r_0=1-2\kappa_\mathrm{ex}/\kappa$ and $r_1 = 1-2\kappa_\mathrm{ex}\gamma/(g^2 + \kappa\gamma)$, respectively, where $\kappa=\kappa_\mathrm{ex}+\kappa_\mathrm{in}$.
By tuning the optimal external coupling rate $\kappa_\mathrm{ex}^\mathrm{opt}=\kappa_\mathrm{in} \sqrt{1 + 2C_\mathrm{in}}$ with the internal cooperativity $C_\mathrm{in} = g^2/(2\kappa_\mathrm{in} \gamma)$, these reflectivities are matched in amplitude while preserving a $\pi$ phase shift between atomic states, so that $r_0^\mathrm{opt}=-r_1^\mathrm{opt}$.
In contrast, the H-polarized photon does not couple to the cavity mode and is routed by  mirrors with unit amplitude $r_\text{m}=1$.

With these reflectivities calibrated, we now follow the evolution of an incoming polarization-entangled state $(\ket{\mathrm{HV}}_p+\ket{\mathrm{VH}}_p)/\sqrt{2}$.
In each term ($\ket{\mathrm{HV}}_p$ or $\ket{\mathrm{VH}}_p$), only one photon enters the cavity and acquires the amplitude $\pm r_0^{\mathrm{opt}}$; its partner is reflected with unit amplitude.
Assuming the two remote nodes are identical, the entangled-pair component at each node is attenuated by the same amount of amplitude
$|r_0^\mathrm{opt}|$, preserving their relative phase and balance.
Therefore, the fidelity of mapping the photonic entanglement onto the memories is unity, while the success probability is given by~\cite{Shapiro2024, kikura2025passivequantuminterconnectshighfidelity}
\begin{equation}
\eta_\mathrm{CAPS} =|r_0^\mathrm{opt}|^2= 1 - \frac{2\sqrt{1 + 2C_\mathrm{in}}}{1 + C_\mathrm{in} + \sqrt{1 + 2C_\mathrm{in}}}.
\end{equation}
as shown in Fig.~\ref{fig:duan_kimble}.
If the two nodes are not perfectly matched, unit fidelity can still be recovered by introducing a calibrated attenuation in the H-polarized path to equalize the attenuation, at the cost of an additional reduction in $\eta_\mathrm{CAPS}$~\cite{kikura2025passivequantuminterconnectshighfidelity}.

Another source of error is temporal-mode mismatch between the two polarization paths~\cite{Utsugi2025, kikura2025passivequantuminterconnectshighfidelity}.
It stems from the finite bandwidth of the photon: off-resonant frequency components probe regions of the atom–cavity response where both the reflection amplitude and phase vary with detuning.  
As a result, the V-polarized photon acquires a group delay that depends on the atomic state, $\ket{0}_a$ or $\ket{1}_a$, so the photons at the two paths exit the interferometer at slightly different times.  
We mitigate this by inserting a fixed optical delay in the H-polarized path equal to the average of the two state-dependent delays; any residual time difference can be further suppressed by fine-tuning the cavity length.  
For the parameters relevant here $(C_\mathrm{in}\!\sim\!100$, gate time $\sim\!100\,\text{ns}$), these measures yield fidelities exceeding 99~\%~\cite{kikura2025passivequantuminterconnectshighfidelity}.

\subsubsection{Entanglement swapping}
\label{sec:entswap}
When neighboring quantum repeater stations each establish an entangled pair of atoms, long-distance entanglement across the entire repeater chain can be achieved through entanglement swapping. 
The key requirement for this process is to perform a Bell measurement on one atom from each established pair in a manner that is as close to deterministic as possible. In neutral-atom systems, nearly deterministic Bell measurements can be conducted using two-qubit gates facilitated by the Rydberg interaction, rather than relying solely on probabilistic optical Bell-state measurements. 
This method circumvents the 50\% intrinsic failure probability associated with linear optics Bell measurements, significantly enhancing the throughput of entanglement swapping in quantum repeater architectures based on neutral atoms.

To illustrate the entanglement swapping procedure in Fig. \ref{Figure 1}, we consider two neighboring links, each having generated an entangled pair of atoms, labeled (A1, A2) and (B1, B2). 
By employing a movable optical tweezer array, one atom from each pair—A2 and B1—are brought into proximity to undergo a CZ gate operation via their Rydberg interaction.
In the Rydberg CZ gates, $\ket{1}_a$ is resonantly coupled to a high-lying Rydberg state $\ket{r}_a$ while  $\ket{0}_a$ is unaffected by the Rydberg laser.
Once an atom is excited to $\ket{r}_a$, it strongly shifts the transition energy of any nearby atom within the blockade radius, preventing both from simultaneously occupying $\ket{r}_a$. Consequently, a conditional phase is acquired only in the $\ket{11}_a$ branch of the joint state of two atoms, thereby implementing a CZ gate \cite{Saffman2010}. 
Experimentally, gate fidelities of $F_\mathrm{Ry}=99.5$ \% have been demonstrated with microsecond-scale operation times \cite{Bluvstein2024}, which is rapid compared to typical atomic coherence times. 
Upon completing the CZ gate, local single-qubit rotations and state-selective readouts in the $X$ basis project A2 and B1 into a Bell measurement outcome. 
This outcome heralds the creation of an entangled state between the remaining atoms in the chain, A1 and B2, thereby swapping entanglement over a greater distance. 
The heralded measurement outcome is communicated classically to the relevant nodes, allowing for a local Pauli correction if needed to ensure that the A1 and B2 is converted to a predetermined Bell state such as $\frac{1}{\sqrt{2}}(\ket{00}_a+\ket{11}_a)$. 

Because this swapping process relies on a deterministic Rydberg-gate based Bell measurement, the overall success probability approaches unity. 
The primary sources of infidelity include residual Rydberg gate errors, imperfect readout, and spin decoherence while awaiting additional links. 
As discussed in the Section \ref{sec:rates}, under realistic conditions, achieving an entanglement fidelity above 90\% per swap is readily attainable, allowing for multiple rounds of swapping to produce high-fidelity entangled pairs across continental or even global distances. 
Further details on the impact of spin dephasing as the dominant error source are discussed in Appendix~\ref{app:rfidplots}.

\section{Entanglement distribution rates and fidelities}\label{sec:rates}
Like many terrestrial repeater schemes, this repeater architecture is nested \cite{PhysRevA.91.052325}. The entanglement distribution rate during a flyby is given by
\begin{equation}
R=R_s\eta_sP_0\eta_{\text{CAPS}}\eta_dP_{\text{ES}},  
\end{equation}
where $R_s$ is the photon pair source rate, and $\eta_s$ is the source efficiency. $P_0$ is the average two-photon transmission during the flyby, which is defined as $P_0=\int^{T_{\text{FB}}}_{0}{\eta^{(2)}_{\text{tr}}(t) \,dt}/T_{\text{FB}}$ with $\eta^{(2)}_{\text{tr}}(t)$ being the time-dependent two-photon transmission. $\eta_{\text{CAPS}}$ is the memory loading efficiency, and $\eta_d$ is the detection efficiency. $P_{\text{ES}}=(\frac{2}{3}P_{\text{gate}})^n$ is the swapping probability with $n$ being the number of nesting levels, i.e., $N_{\text{link}}=2^n$. Then, the total distribution distance is $L=2^nL_0$. The gate efficiency depends on the type of gate. For the Rydberg gate, it is 1.   



\begin{figure}
\centering
  \includegraphics[scale=0.58]{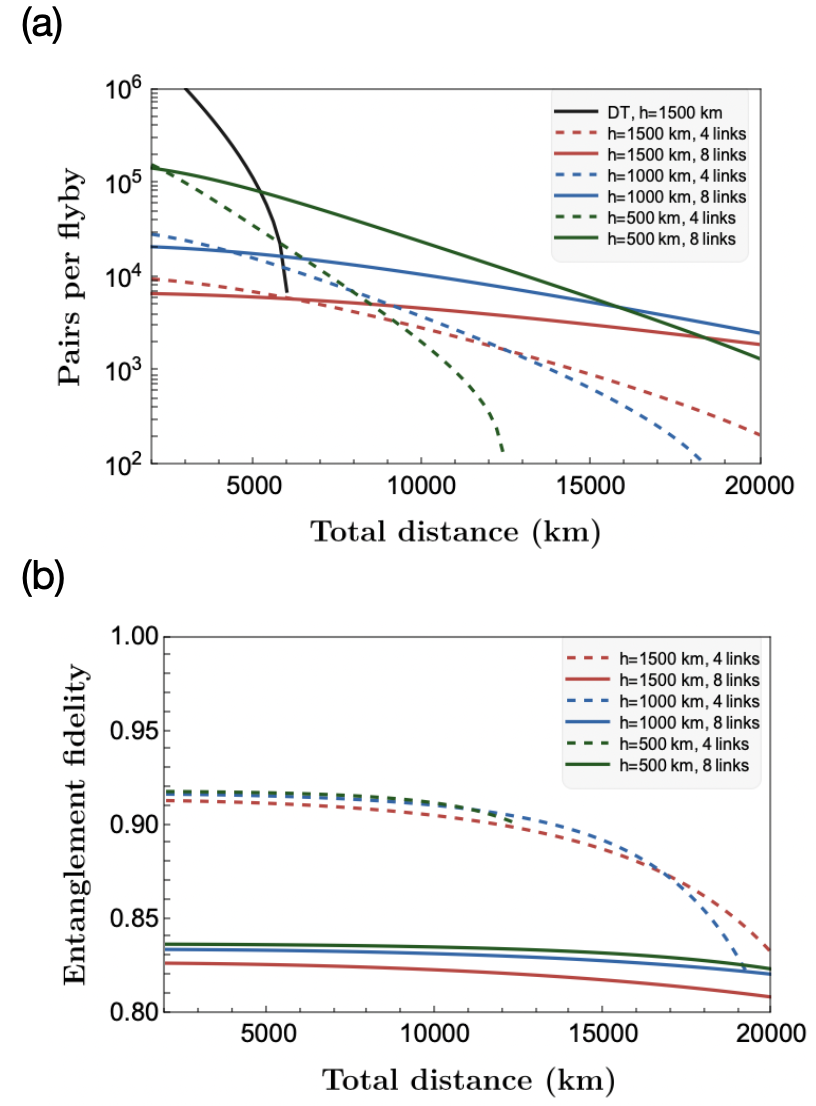}
  \caption{(a) The number of distributed pairs per flyby as a function of total distance for multiplexed quantum repeaters with LEO satellites at different heights and direct transmission (DT) from a single satellite at 1500 km. The dotted lines represent 4-link repeaters, and the solid lines represent 8-link repeaters. (b) The fidelities of distributed pairs as a function of total distance for the repeaters. The relevant parameters used in this figure can be found in the Table. \ref{table: parameters}. }  
  \label{hrates}
  \end{figure}

To enhance the distribution rates, one could implement spatial multiplexing by employing multiple cascaded SPDC sources in parallel on each satellite. In addition, frequency multiplexing can boost channel capacity by emitting photon pairs at different wavelengths. Upon reception of these photon pairs on the ground, their distinct wavelengths must be shifted back to the original wavelength before the photons enter the cavity. The implementation of this mixed multiplexing strategy is discussed in detail in Sec. \ref{sec:implementation}. In general, the distribution rate is given as follows:
\begin{equation}
R_{\text{mux}}=N_{\text{mux}}\eta^2_{\text{demux}}R_s\eta_s\bar{P_ 0}\eta_{\text{CAPS}}\eta_d P_{\text{ES}}, 
\end{equation}
where $N_{\text{mux}}$ is the number of multiplexing channels, $\eta_{\text{demux}}$ is the demultiplexing efficiency, and $\bar{P_0}$ is the mean of the two-photon transmission $P_0$ for all different frequency channels. Since the total shift in wavelength is within a few nm for these distinct channels, this mean value $\bar{P_0}$ is very close to $P_0$. The rates can be converted to the number of pairs generated per flyby, which is given by $N_{\text{FB}}=R_{\text{mux}}T_{\text{FB}}$. For direct transmission, there is only one link with a single satellite, thus $n=0$. The distribution rate is given by
\begin{equation}
R_{\text{dir}}=N_{\text{mux}}R_0\eta_s\bar{P_0}, 
\end{equation}
where $R_0$ is the photon pair source rate for direct transmission.


The fidelity of the final distributed pair is mainly affected by spin decoherence, operation infidelities, and initial photon pair fidelity. Since this architecture is nested, the fidelity of a repeater with $n$ nesting levels is given by
\begin{equation}
F_n=F_{\text{Ry}}\times F^2_m\times(\frac{1}{4}+(F_{n-1}-\frac{1}{4})e^{-\gamma_sT_n})F_{n-1},    
\label{fideq}
\end{equation}
where $\gamma_s$ stands for the spin decoherence rate, $F_m$ the measurement fidelity on the atom, $F_{\text{Ry}}$ the Rydberg gate fidelity, and $T_n$ the average waiting time for the nth level to be established. The third term in Eq.~(\ref{fideq}) characterizes the link degraded by spin decoherence while waiting for the other link to be established. With multiplexing, the waiting time takes the following form:
\begin{equation}
T_n=\frac{\frac{3^{n-1}}{2^n}}{N_s\eta^2_{\text{demux}}R_s\eta_s\bar{P}_0\eta_{\text{CAPS}}\eta_d},  
\end{equation}
where $n\geq 1$. Eq.~(\ref{fideq}) is iterative as it depends on the fidelity of the n-1th level. $F_0$ stands for the Werner state fidelity of the elementary link, which is given by

\begin{equation}
F_0=\frac{4\bar{F}_{\text{pair}}F_{\text{CAPS}}-1}{3},  
\end{equation}
where $F_{\text{CAPS}}$ is the CAPS gate fidelity of memory loading, and $\bar{F}_{\text{pair}}$ is the average pair fidelity during the flyby, which is given by

\begin{equation}
\bar{F}_{\text{pair}}=\frac{1}{P_0T_{\text{FB}}}\int^{T_{\text{FB}}}_{0}{F_{\text{pair}}(t)\eta^{(2)}_{\text{tr}}(t) dt},   
\end{equation}
where $F_{\text{pair}}(t)$ is given in Eq.~(\ref{eq:pairfid}). Here, we neglect the effect of dark counts as the rate can be as low as mHz in superconducting nanowire detectors \cite{Schuck2013}. The detailed derivations of the iterative repeater fidelity and the waiting time can be found in Appendix \ref{app:rfid}.

In Fig.~\ref{hrates}(a), we plot the number of pairs per flyby for the repeaters at various heights and direct transmission from a single satellite at 1500 km. The general trend is that the more links the repeater has, the better its performance. The only 4-link repeater that can reach 20,000 km is one with the satellite at 1,500 km. For 8-link repeaters, the one with the satellite at 1000 km outperforms the others at 20000 km, yielding around 3000 pairs. Fig.~\ref{hrates}(b) presents the fidelities versus the distributed distance. In general, the 4-link repeaters outperform the 8-link repeaters, and the height has a negative impact on the fidelity, with a higher altitude resulting in lower fidelities. At 20000 km, the 4-link repeater with h=1500 km yields the fidelity of around $84\%$, and the corresponding 8-link repeater yields a fidelity of around $81\%$. In Appendix~\ref{app:rfidplots}, we also show the impact of spin decoherence rate and source fidelity on the repeater fidelity by comparing two values they could take.


\section{Implementation}\label{sec:implementation}


\begin{table*}
\centering
\begin{tabular}{|c c c|} 
\hline
Parameter list &Symbol & Value\\
\hline
Source repetition rate for repeaters & $R_s$ & 10 MHz \\ 
Source repetition rate for direct transmission & $R_0$ & 1 GHz \\ 
Source emission efficiency & $\eta_s$ & 0.9\\
Source pair fidelity & $F_s$ & 0.998\\
Height of satellite & h & 1500 km \\
Receiver radius & $D_{\text{R}}$ & 1 m \\
Beam quality factor & $M^2$ & 1  \\
Initial beam waist & $\omega_0$ & 2.5 cm \\
Standard deviation of pointing error & $\sigma_{\text{point}}$ & 0.5 $\mu$rad \\
Photon wavelength & $\lambda$ & 780 nm \\
Transmittance at zenith & $\eta_z$ & 0.79 \\
Detection efficiency & $\eta_d$ & 0.9 \\
System coupling efficiency & $\eta_{\text{cpl}}$ & 0.25 \\
Frequency demultiplexing efficiency & $\eta_{\text{demux}}$ & 0.73 \\
Rydberg gate fidelity & $F_{\text{Ry}}$ & 0.995 \\
The CAPS gate efficiency &$\eta_{\text{CAPS}}$ & 0.75 \\
The CAPS gate fidelity &$F_{\text{CAPS}}$ & 0.99 \\
Atom readout fidelity & $F_m$ & 0.999\\
Spin decoherence rate & $\gamma_s$ & 50 mHz\\
Frequnecy multiplexing channels & $N_{\text{mux}}$ & 100\\
Maximum zenith angle &$\theta_m$ & 80\degree \\
Field of view (FOV) &$\Omega_{\text{fov}}$ & 100 $\mu$rad \\
Spectral irradiance &$H$ & $1.5\times 10^{-5}$ Wm$^{-2}$$\mu$m$^{-1}$sr$^{-1}$ \\
Filter bandwidth &$\Delta\lambda$ & 1 nm \\
Coincidence time window &$\Delta T$ & 1 ns \\
\hline
\end{tabular}
\caption{The assumed parameter values in the model. The values are consistent with the capabilities of existing or near-term technology. The detailed experimental feasibility of these values is discussed in Sec. \ref{sec:implementation}. The parameters that are different from the ones shown in the table are indicated in the figures. }
\label{table: parameters}
\end{table*}

\begin{figure*}
\centering
  \includegraphics[width=0.9\textwidth]{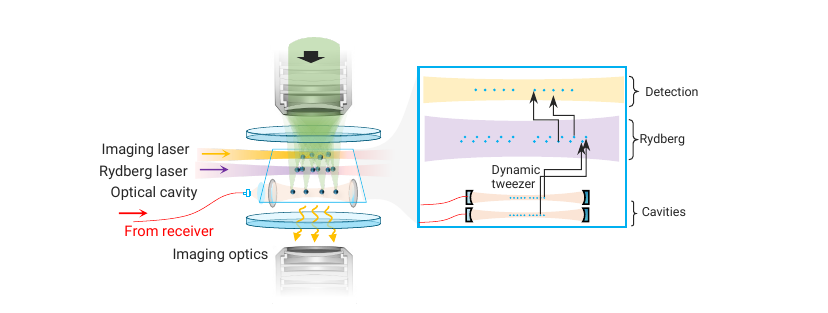}
  \caption{Overview of a neutral-atom quantum processing unit (QPU) module interfaced with optical cavities. The QPU features a high-numerical-aperture (high-NA) objective lens, which focuses on a plane where single atoms are trapped. 
  Static and dynamic optical tweezers, generated by spatial light modulators (SLMs) and acousto-optic deflectors (AODs), are focused through the objective to create tight atom traps. Dichroic mirrors combine multiple control wavelengths—including those for imaging, Rydberg excitation, and cavity coupling (hiding laser)—allowing all beams to pass through the same lens. 
  Fluorescence emitted by atoms is collected through imaging optics for qubit state readout.
  The inset illustrates a zoned architecture where atom arrays are dynamically shuttled between three primary regions using optical tweezers. Once entangled atom pairs are heralded by photon detection through the CAPS gates in each cavity, entanglement swapping is carried out as follows: the atoms are transported to the Rydberg interaction zone, where a Rydberg CZ  gate is applied between the atom pair. The atoms are then moved to the detection region, where an $X$-basis measurement projects the system into an entangled state between distant atoms, thereby extending entanglement across the network. 
  }
  \label{fig:qpu_impl}
  \end{figure*}

This section discusses the practical implementation of the satellite-assisted quantum repeater architecture described in Section \ref{sec:quantum archi}. 
We detail the technical realization of each component, including the satellite-based photon-pair source, atom-photon entanglement mapping using the CAPS gate, and neutral-atom QPUs. 
These components are characterized by a feasible set of system parameters, summarized in Table \ref{table: parameters}, which support viable entanglement distribution rates and fidelities over intercontinental distances, far surpassing the capabilities of terrestrial quantum repeaters.




\subsection{Satellite-Borne Entangled Photon Pair Source}
\label{sec:photonsource}
Various types of entangled photon pair sources have been developed, including quantum dots \cite{Basset2021, Zahidy2024, Yang2024}, color centers \cite{Rahmouni2024}, and atomic vapor \cite{Craddock2024}, each offering unique advantages in terms of efficiency, coherence, and scalability. 
Despite these advancements, spontaneous parametric down-conversion (SPDC) remains the only method successfully implemented in satellite-based experiments, establishing it as the most viable option for space-borne entangled photon sources.
For instance, the Micius quantum satellite has already demonstrated the capability of SPDC-based entanglement distribution, achieving a ground-to-ground entanglement link that spans over 1100 km, with a secure key rate of 0.12 bit/s \cite{Yin2020}. 
While this achievement highlights the potential of space-borne entangled photon sources, there is a critical need to enhance their rate to support long-distance quantum repeater networks.
Recent advancements in on-chip SPDC photon-pair generation, i.e., silicon nitride \cite{Alexander2025} and lithium niobate/tantalate waveguides and cavities \cite{Zhao2020, Wang2024}, offer promising avenues for compact, power-efficient, and scalable photon-pair sources, making them ideally suited for satellite deployment and the development of large-scale quantum networks.

However, the inherently probabilistic nature of SPDC poses a significant challenge for high-rate entanglement generation, as each photon pair is produced with a limited success probability to mitigate multiphoton contamination. 
To address this challenge, spatial and/or frequency multiplexing protocols play a key role in enhancing the photon-pair generation rate. 
Spatial multiplexing employs multiple cascaded SPDC sources in parallel, each undergoing a linear-optical Bell-state measurement between photons belonging to different pairs to herald high-fidelity entangled photon pairs. 
The output from these sources is routed through a photonic switching array, which selects successful entanglement events and directs them into the network, increasing the probability of generating an entangled pair per operational cycle \cite{Dhara2022}. 
While the scalability of spatial multiplexing is currently limited by optical loss in switching arrays, rapid progress in ultra-low loss active photonic circuits, driven by development of scalable photonic quantum computing platforms \cite{Koen2024, AghaeeRad2025}, could offer a practical route towards achieving the parameter regime listed in Table \ref{table: parameters}.

Another critical approach is frequency multiplexing, which boosts channel capacity by simultaneously generating multiple entangled photon pairs at distinct frequencies within a broad phase-matching bandwidth of about 10\,THz~\cite{Chen2022}. 
By exploiting this wide bandwidth in nonlinear crystals, hundreds of frequency channels ($N_\mathrm{mux}\gtrsim 100$) can be populated, dispensing with the need for active switching arrays and thereby reducing the satellite’s overall system complexity. 
Moreover, the heralding signal from the source includes a ``frequency tag,'' allowing the receiving station to identify and dynamically shift each photon to the corresponding atomic or cavity resonance. 
With these capabilities, frequency-multiplexed sources can reach entanglement-generation rates on the order of $\sim 10$\,GHz—without spectral-bandwidth matching, which is addressed in the next paragraph—while preserving an initial pair fidelity of 
$F_s \approx 99.8 \%$~\cite{Chen2022}.

A remaining technical challenge lies in matching the spectral bandwidth of entangled photon pairs to that of the atom-cavity system, ensuring efficient photon-to-atom entanglement transfer through the CAPS gate using optical cavities. 
Typically, SPDC sources exhibit spectral bandwidths that are several orders of magnitude broader than the 10–100 MHz range of an atom–cavity system.
To overcome this mismatch, cavity-enhanced SPDC can provide high spectral brightness within a narrow bandwidth, thereby improving spectral compatibility with atom-cavity system.
Notably, cavity-enhanced SPDC systems designed for alkaline atoms have demonstrated photon pair generation rates approaching the MHz level \cite{Tian2016, Tsai2018, Prakash2019}.

Leveraging the scalability of frequency multiplexing, cavity-enhanced SPDC sources have the potential to achieve an overall entanglement generation rate of 1 GHz, with the combination of a single channel rate $R_s\sim10$ MHz
and $N_\mathrm{mux}\sim 100$ frequency multiplexed channels.
As outlined in Table~\ref{table: parameters}, such an approach harnesses high spectral brightness and mode-matching within a narrower bandwidth, thereby ensuring compatibility with atomic transitions while significantly boosting the photon-pair generation rate—an essential step for enabling high-speed, long-distance entanglement distribution in satellite-based quantum networks.

\subsection{Optical cavities: Entanglement mapping}
\label{sec:mux_dk}
In this section, we discuss the implementation of photon-to-atom entanglement mapping through the CAPS gate utilizing optical cavities. 
It is essential to optimize the entangled photon-atom mapping rate to align with the multiplexed SPDC source. 
Specifically, we aim for (i) a single-channel rate of approximately \( R_s \sim 10 \) MHz and (ii) around \( N_\mathrm{mux} \sim 100 \) frequency-multiplexed channels.

To achieve a single-channel entanglement rate of approximately \(R_s \sim 10\)\,MHz, we propose leveraging time-multiplexed CAPS gate operations, employing multiple atoms strongly coupled to a cavity. 
This approach parallels high-rate remote entanglement generation via photonic Bell-state measurements at the midpoint~\cite{Li2024, sunami2024scalablenetworkingneutralatomqubits}, where reducing overhead for atom preparation and transport enhances the overall entanglement rate. 
Notably, the rate is primarily constrained by the bandwidth of the atom-cavity response, as discussed in Section~\ref{sec:entmap}~\cite{kikura2025passivequantuminterconnectshighfidelity}.

Recent experiments have demonstrated strong coupling of multiple atoms in tweezer arrays to both Fabry--P\'erot~\cite{Yan2023, Liu2023, Grinkemeyer2024} and running-wave cavities~\cite{Peters2024}, attaining coupling rates \(g\) in the tens to hundreds of MHz. Furthermore, time-multiplexed atom--photon entanglement with up to six atoms has shown that success probabilities grow with increasing atom numbers~\cite{Hartung2024}.
Scaling the number of atoms coupled to the cavity to several hundred is thus expected to boost the entanglement mapping rate per channel to the limits imposed by the atom--cavity interaction --- potentially exceeding 10\,MHz. For instance, with \(g/2\pi \approx 30\)\,MHz and \(\kappa/2\pi \approx 10\)\,MHz, a rate above 10\,MHz is well within reach. Consequently, the mapping rate that can be realized with practical cavity parameters comfortably meets (and could exceed) the 10\,MHz benchmark of the SPDC photon-pair source.

To interface frequency-multiplexed photon pairs with $N_\mathrm{mux} = 100$ using multiple atom--cavity systems, each frequency channel must first be demultiplexed and selectively shifted into the atom--cavity resonance.
Specifically, each photon pair is ``tagged'' at the satellite station by its heralding signal, which identifies the pair's frequency. 
This frequency information is forwarded via a classical channel to the ground station, where dynamic frequency conversion is applied so that each photon matches the atomic transition. 
Advanced waveguide-based nonlinear crystals demonstrate the capability to achieve fiber-to-fiber conversion efficiencies of $\eta_\mathrm{demux}$ above $70\%$ over a broad spectral range~\cite{Clark2013, Murakami2023}, while preserving fidelity.

Once the photons are precisely tuned to the atom--cavity resonance, each channel is directed into a corresponding optical cavity. 
By integrating multiple cavities within the imaging system's field of view of a single QPU, CAPS gate operations can be parallelized across numerous spatially distinct resonators.
Compact cavity designs---such as fiber-based Fabry--Pérot cavities~\cite{Brekenfeld2020, Grinkemeyer2024} and micro-mirror arrays~[cite], on-chip photonic cavities~\cite{Zhou2024}, or nanofiber cavities~\cite{sunami2024scalablenetworkingneutralatomqubits}---allow dense packing of tens or even hundreds of individual resonators ($N_\mathrm{s,mux}\sim 10\text{--}100$). 
This high degree of parallelization significantly boosts the total entanglement generation rate, while keeping the overhead of atom preparation and transport at a manageable level.


Another route for interfacing a multiplexed photon source is to use frequency-multiplexed CAPS gates. 
These gates allow multiple frequency channels to be mapped simultaneously onto atoms by taking advantage of multiple cavity resonances, each spaced by an integer multiple of the cavity free spectral range. 
By aligning the frequency intervals of photon pairs with adjacent cavity resonances and applying a light shift to the corresponding atomic transitions, it becomes possible to execute multiple CAPS gates in a single operation across different frequency components~\cite{kikura2025passivequantuminterconnectshighfidelity}.
Each frequency-matched photon then interacts with its designated atom via these light-shift beams, enabling atom--photon entanglement across multiple spectral channels in parallel.
Once the frequency-multiplexed photons have been reflected from the atom--cavity system, they still contain multiple frequency components, requiring frequency demultiplexing for individual channel heralding.
This can be efficiently achieved using highly dispersive optical elements, including high-finesse cavities, diffraction gratings, or tunable optical filters, thereby enabling precise detection and processing of each frequency component.
With a light-shift capability of roughly 10\,GHz~\cite{Hu2024} and a free spectral range of about 1\,GHz for a 10-cm-long cavity~\cite{Kato2019}, a multiplexing factor of $N_\mathrm{f,mux}\sim 10$ is within reach using current technology.

Consequently, combining spatial multiplexing ($N_\mathrm{s,mux} \sim 10$) and frequency multiplexing ($N_\mathrm{f,mux} \sim 10$) in CAPS gates---each with a single-channel capacity of around 10\,MHz---can collectively reach the multiplexing number of 100. As indicated in Table~\ref{table: parameters}, this comfortably matches the total rate of the photon-pair source, thereby providing the high remote entanglement throughput required for a terrestrial quantum repeater network.


\subsection{Neutral-atom QPUs: Entanglement swapping}


In this section, we present neutral-atom QPUs that facilitate the entanglement swapping operation outlined in Section~\ref{sec:entswap}. 
One of the key features of neutral atom QPUs is their capacity to physically transport atomic qubits across various zones while preserving coherence \cite{Bluvstein2022}, using dynamically controlled optical tweezer arrays implemented through acousto-optic deflectors, as illustrated in  Fig.~\ref{fig:qpu_impl}(a).
This zone architecture leverages atom transport to physically separate a Rydberg zone for two-qubit gates from the cavity and imaging zones, ensuring that qubits not actively engaged in Rydberg operations or readout remain isolated—thereby mitigating crosstalk and maintaining coherence. Since typical transport times are on the order of milliseconds---much shorter than the second-scale coherence time of the atoms---the error associated with transport is negligible \cite{Bluvstein2024}.  
Such functionality not only offers flexible connectivity but also proves particularly advantageous for large-scale quantum computing.

After entangled atom pairs are heralded in each cavity by CAPS gates, one atom from each pair is transported to the Rydberg zone to execute a Rydberg-based CZ operation~\cite{Saffman2010}. 
In this process, atoms are excited from one of the qubit states to highly excited Rydberg states, where the presence of an already-excited Rydberg atom shifts the energy levels of its neighboring atoms due to the strong dipole-dipole interaction.
This interaction prevents further excitation within the blockade radius, enabling a conditional phase shift on the atomic qubits and facilitating high-fidelity entangling gates.
The CZ gate fidelities of Rb and other atomic species reach or even exceed $F_\mathrm{Ry}=99.5$ \% within a typical timescale of sub $\mu$s using time-optimal pulse sequences to minimize the effect of Rydberg decay ~\cite{Evered2023,Peper2025}.

Upon completion of the CZ gate, the atoms are moved to the detection region, where an $X$-basis measurement projects the system into an entangled state across atoms in neighboring nodes.
By default, the hardware measures the $Z$ basis, detecting fluorescence from one of the atomic qubit states—a process that takes a few milliseconds and achieves a readout fidelity above 99.9\% \cite{Chow2023,Manetsch2024}.
To realize an $X$-basis measurement, a Hadamard gate is first applied to rotate the qubit states from the $X$ basis into the $Z$ basis. 
These single-qubit operations can be performed via Raman coupling \cite{Ma2022, Jenkins2022, Levine2022} with fidelity exceeding 99.9\% at a gate speed of about one microsecond. 
Global rotations are achieved by illuminating all or part of the atom array with large laser beams, whereas individual atoms can be addressed using tightly focused beams. 
This yields an overall X-basis measurement fidelity of $F_m=99.9$\%. 
By repeating this entanglement-swapping cycle across multiple nodes, long-distance quantum entanglement can be systematically extended.

\section{conclusions and outlook}
\label{sec:conclusion}
We propose a global quantum network architecture based on neutral atoms in optical cavities and LEO satellite links. A central feature of this architecture is spatial and frequency multiplexing, which significantly boosts the number of distributed entangled pairs and underpins the repeater’s overall performance. Under realistic assumptions, our analysis shows that a four-link repeater can distribute approximately 300 entangled pairs over 20,000 km during a satellite flyby, achieving a fidelity close to $85\%$. With eight links and multiplexing, the system can distribute around 3,000 pairs per flyby over the same distance, maintaining a fidelity above $80\%$. We further investigate how key parameters—including satellite altitude, source fidelity, and spin decoherence rate—impact the performance of the repeater architecture. Finally, we discuss in detail the implementation of the proposed architecture in the cavity-based neutral atom systems.

Here, we have focused on the quantum network with ground stations on the equator and the LEO satellites in polar orbits. Looking ahead, it would be valuable to extend this study to scenarios involving a global network of ground stations, each equipped with atomic memories, distributed more broadly across the Earth’s surface. Such a network could interface with a constellation of satellites, as proposed in \cite{Khatri2021}, enabling the estimation of the number of entangled pairs that could be distributed per day across diverse geographic locations under more realistic operational conditions. This would provide critical insights into how factors such as orbital coverage, ground-station density, and regional weather patterns influence network performance.  

Furthermore, one could consider a more realistic situation where ground stations located in the suburbs need to communicate with the users in the city, thus requiring terrestrial quantum repeaters for this connection \cite{Simon2017}. Neutral atoms present a promising platform for building this global hybrid quantum network, offering the potential for multiplexed quantum networking at telecom wavelengths on the ground \cite{PhysRevResearch.3.043154}.        

\section*{Declaration of Competing Interest}
This work was supported by Nanofiber Quantum Technologies, Inc., under a sponsored research agreement with the University of Calgary. S. Sunami and S. Kikura are employees, and A. Goban is a co-founder and a shareholder of Nanofiber Quantum Technologies, Inc.


\appendix

\section{SATELLITE DYNAMICS}
\label{app:dynamics}

\begin{figure}
\centering
  \includegraphics[scale=0.5]{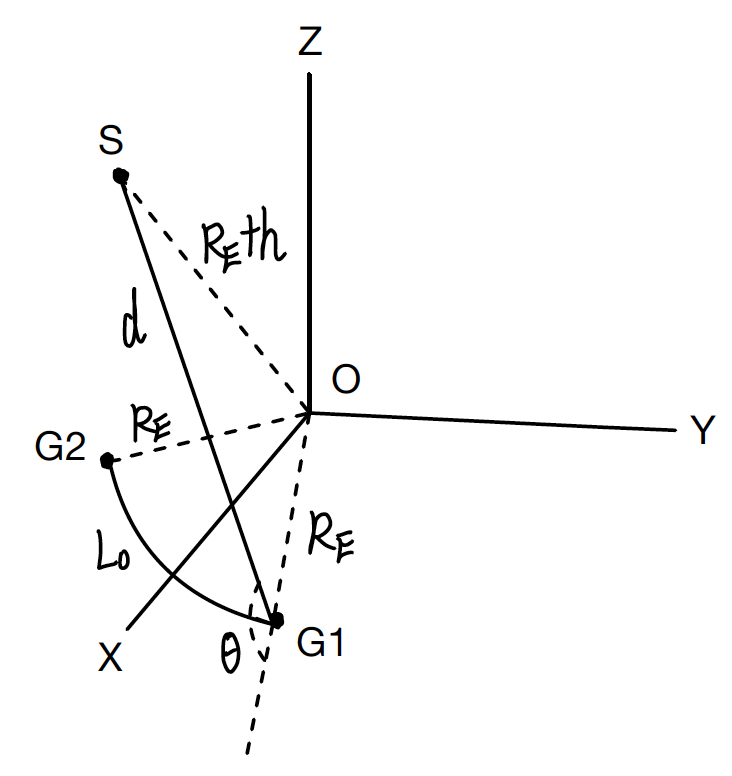}
  \caption{The schematic of the geometric relations of the satellite and two ground stations. Point S stands for the satellite, which moves in a polar orbit in the X-Z plane, and G1 and G2 stand for two ground stations on the equator in the X-Y plane. $\theta$ is the zenith angle. The distance between the ground stations is $L_0$, the elementary link length.}  
  \label{cosys}
  \end{figure}

The satellite is in a polar orbit, and the ground stations are on the equator. In the Cartesian coordinate system, they can be written as follows:

\begin{equation}
\begin{aligned}
\text{Ground station}: &\left(R_E\cos\frac{L_0}{2R_E}, R_E\sin\frac{L_0}{2R_E}, 0\right),\\
\text{Satellite}: &\Big((R_E+h)\cos\omega (t_0-t), 0,\\ &(R_E+h)\sin\omega (t_0-t)\Big),  
\end{aligned}    
\end{equation}
where $L_0$ is the elementary link length, $R_E$ the earth's radius, $h$ the satellite height, and $\omega$ the satellite angular speed. The satellite starts at a position where the zenith angle is at a maximum. The time $t_0$ is the time that it takes the satellites to fly from the initial position (maximum zenith angle) to the midpoint. Then, the distance can be expressed as:
\begin{equation}
\begin{aligned}
d^2=&\left(R_E\cos\frac{L_0}{2R_E}-(R_E+h)\cos\omega (t_0-t)\right)^2\\
&+\left(R_E\sin\frac{L_0}{2R_E}-0\right)^2+(0-(R_E+h)\sin\omega (t_0-t))^2.
\end{aligned}    
\end{equation}
This can be further simplified as: 
\begin{equation}
\begin{aligned}
d^2=&R^2_E+(R_E+h)^2\\
-&2R_E (R_E+h)\cos\frac{L_0}{2R_E}\cos\omega(t_ 0-t),
\end{aligned}
\end{equation}
which is the same as Eq.~(\ref{eq:dis}).

To compute the time $t_0$, we assume that the maximum zenith angle is $\theta_m$ at $t=0$. Eq.~(\ref{eq:zenith}) links the zenith angle $\theta$ to the distance $d$, which can be derived using the law of cosines in the triangle SG1O:

\begin{equation}
(R_E+h)^2=d^2+R^2_E-2dR_E\cos(\pi-\theta),   
\end{equation}
which can be further simplified as:
\begin{equation}
h^2+2hR_E=d^2+2dR_E\cos\theta.   
\end{equation}
Then, we rearrange this equation to express $\cos\theta$ as Eq.~(\ref{eq:zenith}). Thus, using this relation, the distance can be written as:
\begin{equation}
d=\sqrt{h^2+2hR_E+R^2_E\cos^2\theta_m}-R_E\cos\theta_m.    
\end{equation}
Now, taking this equation into Eq.~(\ref{eq:dis}) with $t=0$, we obtain:
\begin{equation}
\begin{aligned}
\cos\frac{L_0}{2R_E}\cos\omega t_0&=\frac{R^2_E(1-\cos^2\theta_m)}{h+R_E}\\
&+\frac{\cos\theta_m\sqrt{h^2+2hR_E+R^2_E\cos^2\theta_m}}{h+R_E}.    
\end{aligned}  
\end{equation}
This can be further simplified to find the time $t_0$, which is given by:

\begin{equation}
t_ 0=\frac{\arccos{\frac{R_E(1-\cos^2\theta_m)+\cos\theta_m\sqrt{h^2+2hR_E+R^2_E\cos^2\theta_m}}{(h+R_E)\cos\frac{L_ 0}{2R_E}}}}{\omega}.
\end{equation}

\section{REPEATER FIDELITY AND WAITING TIME}
\label{app:rfid}
The repeater can be seen as two main segments regardless of the nesting level. For the repeaters with $n$ nesting levels, its fidelity is the product of the fidelity of the two sublevels ($n-1$th level), the Rydberg gate fidelity, and the measurement fidelity, which is given by
\begin{equation}
F_n=F_{\text{Ry}}\times F^2_m\times F^{(e)}_{n-1}\times F^{(l)}_{n-1},
\label{B1}
\end{equation}
where the superscripts $e$ and $l$ stand for the early and late. The early segment is the segment that establishes the entanglement first, and the late segment is the segment that establishes the entanglement second. These two fidelities are related to each other through the average waiting time, i.e., the time difference between the completion of the two segments. During this time, the early link experiences spin decoherence, thus degrading its fidelity. Here, we neglect the spin decoherence during the Rydberg gate operation and measurement.

Upon the time when the link is established, its density matrix can be written as the Werner state:
\begin{equation}
\rho_{n-1}=F_{n-1}\ket{\psi^-}\bra{\psi^-}+\frac{1-F_{n-1}}{4}I,    
\end{equation}
where $\ket{\psi^-}=\frac{1}{\sqrt{2}}(\ket{01}-\ket{10})$ is a Bell state. Then, in the basis $\{\ket{00},\ket{01},\ket{10},\ket{11}\}$, this state can be expressed as:
\begin{equation}
\renewcommand\arraystretch{1.2}
\rho_{n-1}=\begin{pmatrix}
 \frac{1-F_{n-1}}{4}  &0   &0 &0 \\
   0     &\frac{1+F_{n-1}}{4} &-\frac{F_{n-1}}{2}  &0\\
 0   &-\frac{F_{n-1}}{2} &\frac{1+F_{n-1}}{4} &0 \\
  0     &0     &0       &\frac{1-F_{n-1}}{4}\\
\end{pmatrix}.
\end{equation}
  
Under the spin decay and spin dephasing, the density matrix becomes:
\begin{widetext}
\begin{equation}
\renewcommand\arraystretch{1.5}
\rho_{n-1}(t)=\begin{pmatrix}
 \frac{1-F_{n-1}}{4}+(1-e^{-\gamma t})\frac{1+F_{n-1}}{4}  &0   &0 &0 \\
   0     &e^{-\gamma t}\frac{1+F_{n-1}}{4} &-\frac{F_{n-1}}{2}e^{-\gamma_s t}  &0\\
 0   &-\frac{F_{n-1}}{2}e^{-\gamma_s t} &e^{-\gamma t}\frac{1+F_{n-1}}{4} &0 \\
  0     &0     &0       &\frac{1-F_{n-1}}{4}e^{-2\gamma t}\\
\end{pmatrix},
\end{equation}
\end{widetext}
where $\gamma$ is the spin decay rate and $\gamma_s=\gamma+2\gamma^*$ is the spin decoherence rate with $\gamma^*$ being the pure spin dephasing rate. Thus, the fidelity with respect to the initial Werner state is given by: 
\begin{equation}
F_{n-1}(t)=\frac{1}{4}+\left(F_{n-1}-\frac{1}{4}\right)e^{-\gamma_s t},    
\end{equation}
and we obtain
\begin{equation}
\begin{aligned}
&F^{(l)}_{n-1}=F_{n-1}(0)=F_{n-1},\\ 
&F^{(e)}_{n-1}=F_{n-1}(T_n)=\frac{1}{4}+\left(F_{n-1}-\frac{1}{4}\right)e^{-\gamma_s T_n},
\end{aligned}    
\end{equation}
where $T_n$ is the average waiting time for the $n$th level to be established. Plugging these relations back into Eq.~(\ref{B1}), we get
\begin{equation}
F_n=F_{\text{Ry}}\times F^2_m\times\left\{\frac{1}{4}+\left(F_{n-1}-\frac{1}{4}\right)e^{-\gamma_sT_n}\right\}F_{n-1}.    
\end{equation}

The waiting time $T_n$ can be computed using the time for establishing the $n-1$th nesting level, which is $(\frac{3}{2})^{n-1}T_0$ with $T_0$ being the average time for establishing an elementary link. This time is given by:
\begin{equation}
T_0=\frac{1}{N_{\text{mux}}\eta^2_{\text{demux}}R_s\eta_s\bar{P}_0\eta_{\text{CAPS}}\eta^2_d}.    
\end{equation}
It is well known that the average waiting time is half of the time for establishing the other segment at the same nesting level \cite{RevModPhys.83.33}, thus we obtain:
\begin{equation}
T_n=\frac{1}{2}\left(\frac{3}{2}\right)^{n-1}T_0=\frac{\frac{3^{n-1}}{2^n}}{N_{\text{mux}}\eta^2_{\text{demux}}R_s\eta_s\bar{P}_0\eta_{\text{CAPS}}\eta^2_d}. 
\end{equation}

\section{THE IMPACT OF SOURCE FIDELITY AND SPIN DECOHERENCE RATE ON THE REPEATER FIDELITY}
\label{app:rfidplots}

\begin{figure}
\centering
  \includegraphics[scale=0.6]{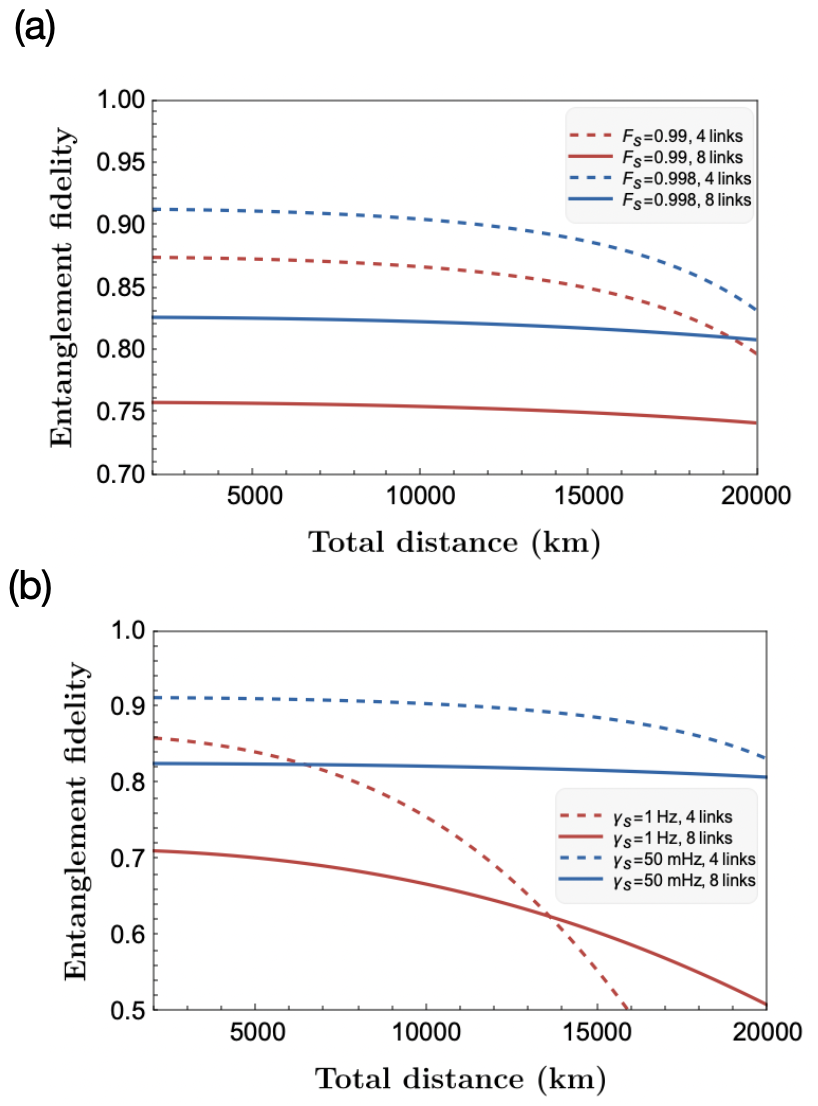}
  \caption{(a) The fidelities of distributed pairs as a function of total distance for the two quantum repeaters with two different source fidelities. The red lines correspond to the repeaters with a source fidelity $F_0=99\%$, and the blue lines correspond to the repeaters with a source fidelity $F_0=99.8\%$. (b) The fidelities of distributed pairs as a function of total distance for the two quantum repeaters with two different spin decoherence rates. The red lines correspond to the repeaters with a rate $\gamma_s=1$ Hz, and the blue lines correspond to the repeaters with a rate $\gamma_s=50$ mHz. The satellite altitude is $1500$ km. The other parameters used in the simulations are presented in the Table. \ref{table: parameters}.} 
  \label{fidelity}
  \end{figure}
  
Fig. \ref{fidelity} illustrates the effect of source pair fidelity and spin decoherence rate on the repeater fidelity for various satellite distances, specifically at $1500$ km. The source fidelity and spin decoherence rate are key factors in determining overall fidelity. Among different configurations, the 8-link repeater is the most sensitive to initial fidelity but the most resilient to spin decoherence. In general, the 8-link repeater has the lowest fidelities due to its largest number of operations and links.

\newpage
\bibliography{mybib}{}

\end{document}